\documentstyle[aps,prl,multicol,floats,epsf]{revtex}

\begin{document}
\draft
\wideabs{

\title{Resistive upper critical fields and irreversibility lines of
optimally-doped high-$T_c$ cuprates}
\author{Yoichi Ando,$^{1,2}$ G. S. Boebinger,$^{1}$ A. Passner,$^{1}$
L.F. Schneemeyer,$^{1}$
T. Kimura,$^{3}$ M. Okuya,$^{3}$ S. Watauchi,$^{3}$
J. Shimoyama,$^{3}$ K. Kishio,$^{3}$
K. Tamasaku,$^{4}$ N. Ichikawa,$^{4}$ and S. Uchida$^{4}$}
\address{$^{\rm 1}$ Bell Laboratories, Lucent Technologies, 
700 Mountain Avenue, Murray Hill, NJ 07974}
\address{$^{\rm 2}$ Central Research Institute of Electric Power
Industry, Komae, Tokyo 201, Japan}
\address{$^{\rm 3}$ Department of Applied Chemistry, University of
Tokyo, Hongo, Bunkyo-ku, Tokyo 113, Japan}
\address{$^{\rm 4}$ Superconductivity Research Course, University of
Tokyo, Yayoi, Bunkyo-ku, Tokyo 113, Japan}

\date{Received Hc2-n1.tex}
\maketitle

\begin{abstract}
We present the resistively-determined upper critical field
$H^{\rho}_{c2} (T)$ and the irreversibility lines $H^{\rho}_{irr} (T)$
of various high-$T_c$ cuprates, deduced from measurements in 61-T
pulsed magnetic fields applied parallel to the $c$ axis. The
{\it shape} of both $H^{\rho}_{c2} (T)$ and $H^{\rho}_{irr} (T)$
depends monotonically on the anisotropy of the material and
none of the samples show saturation of $H^{\rho} (T)$ at low
temperatures. The anomalous positive curvature, $d^2 H^{\rho}/dT^2 > 0$,
is the strongest in materials with the largest normal state anisotropy,
regardless of whether anisotropy is varied by changing the carrier
concentration or by comparing a variety of optimally-doped compounds.
\end{abstract}

}
\narrowtext

The anisotropy of the high $T_c$ cuprates, coupled with their short
coherence lengths, gives rise to a complex magnetic phase diagram,
featuring a vortex liquid state located between the vortex solid and 
the normal state \cite{Fisher2huse,Safar2,Farrell}.
In particular, in the high $T_c$ cuprates, the vortex lattice can
melt at temperatures well below the onset of short-range
superconducting order occurring at the upper critical
magnetic field $H_{c2}$. In a fixed temperature experiment,
this is evidenced by the onset of resistivity due to vortex motion
well below the magnetic field at which the normal-state 
resistivity is restored, defined as the resistively-determined
upper critical magnetic field, $H^{\rho}_{c2} (T)$.  Resistivity
measurements on overdoped ${\rm Tl_2Ba_2CuO_y}$ (Tl-2201)
\cite{Mackenzie}
and overdoped ${\rm Bi_2Sr_2CuO_y}$ (Bi-2201) \cite{Osofsky} have found
a strong upward curvature in $H^{\rho}_{c2}(T)$ with no evidence of
saturation at low temperatures. Such an $H^{\rho}_{c2}(T)$ curve
contrasts strongly with the conventional Werthamer-Helfand-Hohenberg
(WHH) theory for superconductors with weak electron-phonon
coupling \cite{WHH}, in which $H_{c2} (T)$
exhibits negative curvature and saturates at low temperatures.
Similarly anomalous $H^{\rho}_{c2}(T)$ curves have been discussed 
in connection with other anisotropic
superconductors, such as the organic superconductors \cite{Brandow}.
However, due to the extremely large upper critical
fields, studies of the resistive transition in the cuprates have been
largely limited to samples in which $T_c$ and $H_{c2}$ are greatly
suppressed \cite{LosAlamos},
either in strongly underdoped \cite{Karpinska} or overdoped
samples \cite{Mackenzie,Osofsky},
in deliberately impurity-doped samples \cite{Walker}, or in the
electron-doped cuprates \cite{Dalichaouch}.

In addition to the unusual \lq shape' of the
$H^{\rho}_{c2}(T)$ curves, a re-examination of the origin and meaning
of the resistive transition in the cuprates is fueled, in part, by
specific heat \cite{Carrington}, Raman spectroscopy \cite{Blumberg} 
and magnetization \cite{Bergemann} measurements which indicate that
the onset of local superconducting order [generally
interpreted as the mean-field $H_{c2}(T)$] 
in Tl-2201 occurs at magnetic fields well above
$H^{\rho}_{c2}(T)$.
At this mean-field $H_{c2}(T)$, no feature is observed in the measured
resistivity.
Perhaps more surprisingly, the experimental evidence for this 
mean-field $H_{c2}(T)$, in both the specific heat and Raman experiments, 
is anomalously and dramatically suppressed by magnetic fields of only 
a few tesla. Thus, $H^{\rho}_{c2}(T)$ does not seem to correspond to 
the mean-field $H_{c2}(T)$,
which might suggest that it corresponds to
the melting of the vortex lattice. The difficulty with this
interpretation of $H^{\rho}_{c2}(T)$ is that it is unclear why the
melted vortex state should be (or can be) indistinguishable in 
resistivity measurements from the normal state.
More specifically, if $H^{\rho}_{c2}(T)$ corresponds to the
irreversibility line, then
the magnetic phase diagram (at least for Tl-2201 and Bi-2201) contains
an unusual vortex liquid state whose resistivity apparently equals the
normal-state resistivity and exhibits no substantial
magnetic-field or temperature dependence
\cite{Mackenzie,Osofsky}.
In light of the debate about the magnetic phase diagram and the
interpretation of $H^{\rho}_{c2}(T)$ in the cuprates, a systematic study
of $H^{\rho}_{c2}(T)$, particularly in optimally-doped compounds, is
clearly desirable.

In this paper, we present resistivity measurements in 61-T pulsed
magnetic fields which determine $H^{\rho}_{c2}(T)$ for
${\rm La_{2-x}Sr_xCuO_4}$ (LSCO) of various carrier concentrations $x$.
In addition to LSCO, we present $H^{\rho}_{c2}(T)$ data for
nearly optimally doped
${\rm Bi_2Sr_2CaCu_2O_y}$ (Bi-2212) and 
${\rm YBa_2Cu_3O_{7-\delta}}$ (YBCO). We find that
the \lq shape' of the $H^{\rho}_{c2}(T)$ changes monotonically with the
anisotropy of the normal state of the sample, becoming more conventional
as the anisotropy gets smaller. The magnitude of the positive curvature
in $H^{\rho}_{c2}(T)$, along with the steep slope of $H^{\rho}_{c2}(T)$
at the lowest experimental temperature, 
are the greatest in the most anisotropic compounds.

The LSCO crystals, grown by the traveling-solvent floating-zone
method by the University of Tokyo groups, have been previously studied
in pulsed magnetic fields \cite{x-dep-PRL}; the samples reported here
with Sr concentrations $x$=0.08 and 0.17 are grown
by the Kishio group and those with $x$=0.15 are grown by the
Uchida group. 
The Bi-2212 crystals, grown by the Kishio
group using the floating-zone method, are heat-treated in sealed quartz
tubes to tune to the near-optimum doping.
Finally, the YBCO crystals are grown at Bell Laboratories
by a flux method in zirconia crucibles and annealed in sealed quartz
tubes to tune the oxygen content to optimal doping.
The resistivity of each sample was measured at a given
fixed temperature during the magnetic field pulse using a $\sim$100 kHz
lockin technique \cite{x-dep-PRL,Bi2201-PRL}.
No significant eddy current heating resulted
from the time-varying magnetic field. To avoid additional dissipation
due to
vortex depinning in response to an applied Lorenz force, the current
was applied
parallel to the magnetic field (along the $c$ axis), except in the case
of the YBCO crystal, whose shape and the relatively small electrical
anisotropy precluded this geometry.

\begin{figure}
\epsfxsize=0.95\columnwidth
\centerline{\epsffile{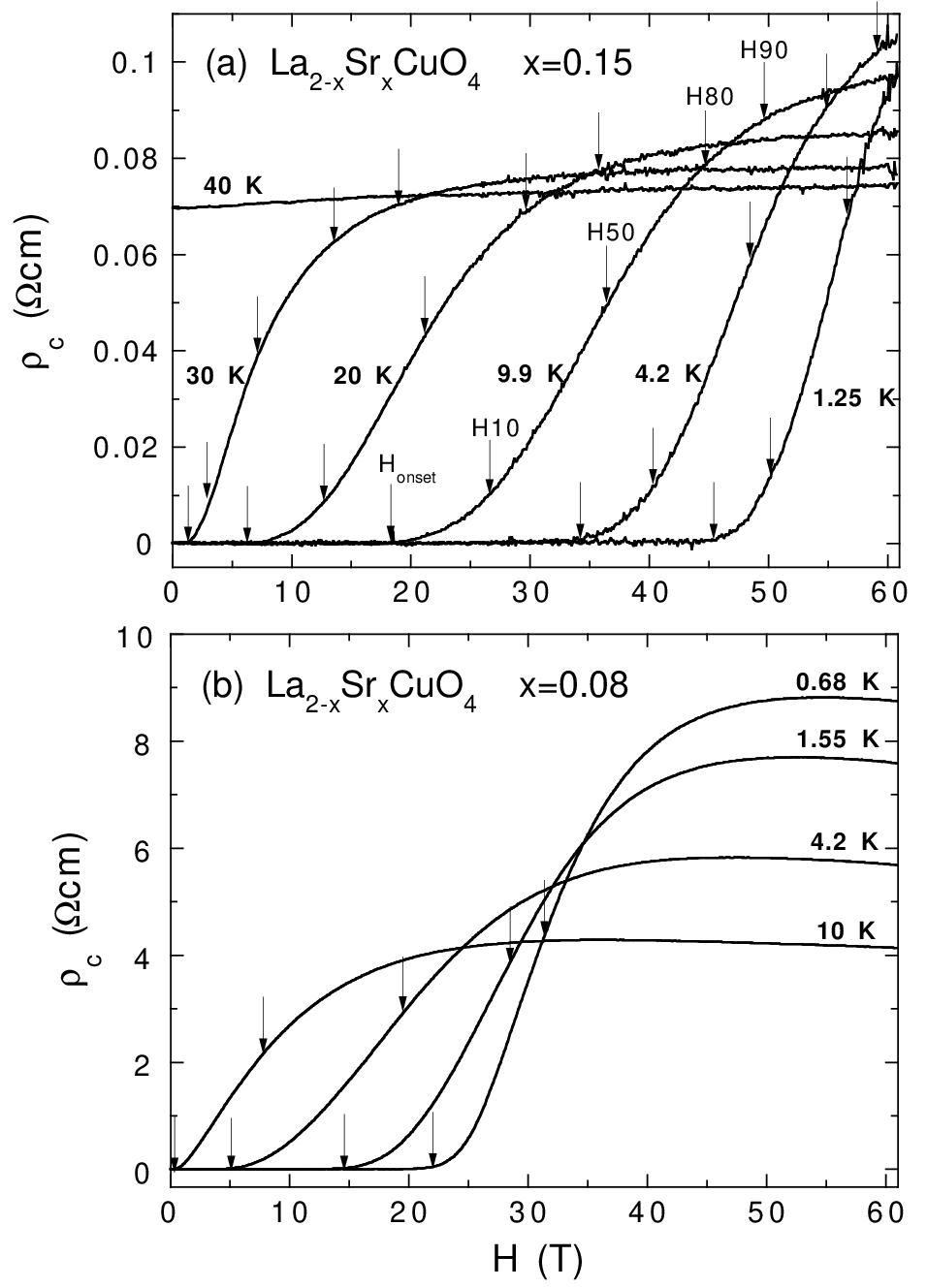}}
\vspace{0.2cm}
\caption{
Selected traces of $\rho_{c}$ vs $H$ of (a) optimally-doped LSCO 
($x$=0.15) and (b) underdoped LSCO ($x$=0.08) at fixed temperatures.  
The arrows in (a), explicitly labelled for the $T=9.9$ K trace, 
mark the onset, 10\%, 50\%, 80\% and 90\% points of each 
resistive transition.  Only $H_{onset}$ and $H50$ are marked by arrows
in (b).}
\label{fig1}
\end{figure}

Figure 1(a) shows a representative set of raw experimental traces,
the $c$-axis resistivity $\rho_{c}$ vs $H$ of LSCO ($x$=0.15)
at various temperatures. The onset field $H_{onset}$ is defined
as the magnetic field at which the resistivity
first is detected to deviate from zero in the $\rho$ vs $H$ plot. This
definition is our best determination of the irreversibility line,
given the limited sensitivity of the pulsed-field data.
To characterize the rest of the resistive transition in the
$H$ vs $T$ plane, we determine those magnetic fields at which the
resistivity equals 10\%, 50\%, 80\%, and 90\% of the ``normal-state"
resistivity,
denoted $H10$, $H50$, $H80$, and $H90$, respectively 
[arrows in Fig. 1(a)].
In traditional superconductors, $H50$ is often associated
with the mean-field upper critical field $H_{c2}$; 
however, given the uncertain
interpretation of the resistive transition in the cuprates, we make
no {\it a priori} determination of $H^{\rho}_{c2}(T)$ from the data.
[We do note, in passing, that $H80(T)$ is comparable to one common
assignment of
$H^{\rho}_{c2}(T)$: the intersection of two straight line 
extrapolations from the normal state resistivity and the steepest 
slope in the transition region.]

At the lowest temperatures, as Fig. 1(a) illustrates, even 61 T
can be insufficient to recover the normal state. In this regime,
$H_{onset}$ remains unambiguous, but the uncertainties in
determining $H10$, $H50$, $H80$, and $H90$ necessarily increase.
In this regime, we extrapolate the normal state resistivity to
those lower temperatures at which it cannot be measured directly.
For the data of Fig. 1(a), this is relatively straightforward since
the low-temperature normal-state resistivity in underdoped LSCO
obeys a $\log (1/T)$ divergence \cite{logT-PRL}; for example, 
we estimate the normal-state resistivity at 1.25 K to be 0.135 
${\rm \Omega}$cm. 
The resulting errors in $H10$ {\it etc.} are estimated to be about 
a few T even at the lowest temperatures. 
For LSCO ($x$=0.08), as shown in Fig. 1(b), no extrapolation
is required and thus the errors in $H10$ {\it etc.} are rather small,
less than 2 T.

\begin{figure}
\epsfxsize=0.95\columnwidth
\centerline{\epsffile{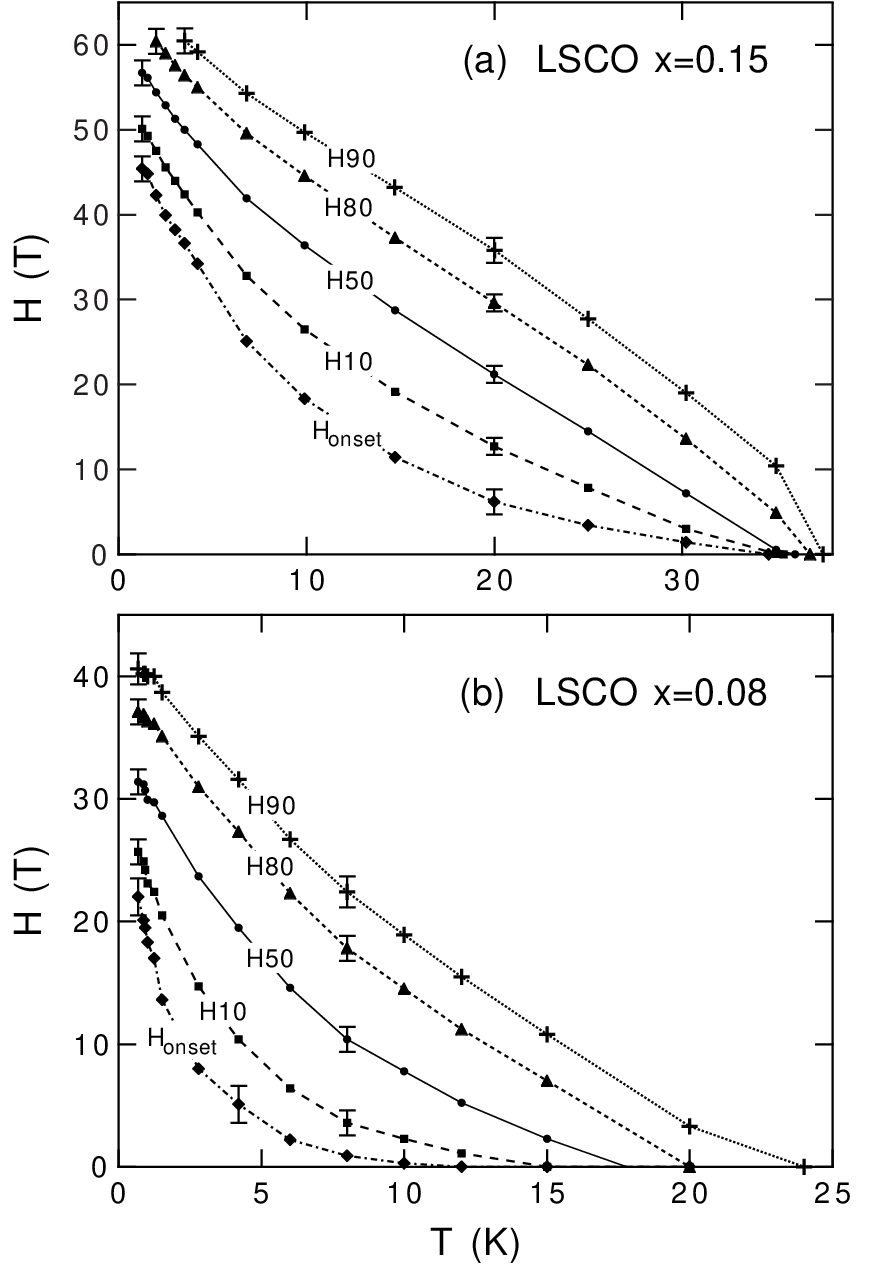}}
\vspace{0.2cm}
\caption{
The resistive transitions for (a) optimally-doped
LSCO ($x$=0.15) and (b) under-doped LSCO ($x$=0.08) in the $H$ vs $T$
plane, characterized by $H_{onset}$, $H10$, $H50$, $H80$, and $H90$.
Representative error bars are attached to the data.}
\label{fig2}
\end{figure}

Figure 2(a) shows the contours of the resistive transition of 
LSCO ($x$=0.15) in the $H$ vs $T$ plane, as determined from the data 
of Fig. 1(a). 
Representative error bars are attached to the data. 
Note that the midpoint of the transition, $H50$, shows a conventional 
linear temperature
dependence from $T_c$ down to $T\simeq 0.5T_c$; however, it exhibits a
steep rise at low temperatures below $\sim 10$ K, $T\simeq 0.25T_c$.
Note also that all of the curves exhibit this sharp rise, which 
persists
to the lowest experimental temperature ($T=1.25$ K $\simeq 0.03T_c$).
Thus, regardless of the precise determination of $H^{\rho}_{c2}(T)$
from the $\rho_c$ vs $H$ curves in Fig. 1(a), 
$H^{\rho}_{c2}$ in optimally-doped LSCO exhibits no evidence of 
saturation at low temperatures, a particularly unusual feature
in common with previously-reported data from overdoped Tl-2201
\cite{Mackenzie} and overdoped Bi-2201 \cite{Osofsky}.

Figure 2(b) shows the contours of the resistive transition for 
underdoped LSCO ($x$=0.08) in the $H$ vs $T$ plane, 
constructed from the data of Fig. 1(b).
Compared to LSCO ($x$=0.15), the curves in Fig. 2(b) show 
even greater upward curvature, 
extending from $T_c$ down to the lowest experimental temperatures
for all curves, including $H80(T)$ and $H90(T)$. 
These data, when coupled with the
data for LSCO ($x$=0.17) [shown later],
suggest a monotonic dependence in which the magnitude of the upward
curvature increases with decreasing carrier concentration.
We emphasize that 61 T is sufficient to suppress superconductivity in 
LSCO ($x$=0.08) and thus all the data points in Fig. 2(b) have rather
small error bars.
Therefore, even though the raw resistive transition in 
Fig. 1(b) is much broader compared to the overdoped Tl-2201 
\cite{Mackenzie} or Bi-2201 \cite{Osofsky}, all the definitions for 
$H_{c2}^{\rho}$ are unambiguously indicating an unusual upward curvature 
of $H_{c2}^{\rho}(T)$ in this underdoped LSCO.

\begin{figure}
\epsfxsize=0.95\columnwidth
\centerline{\epsffile{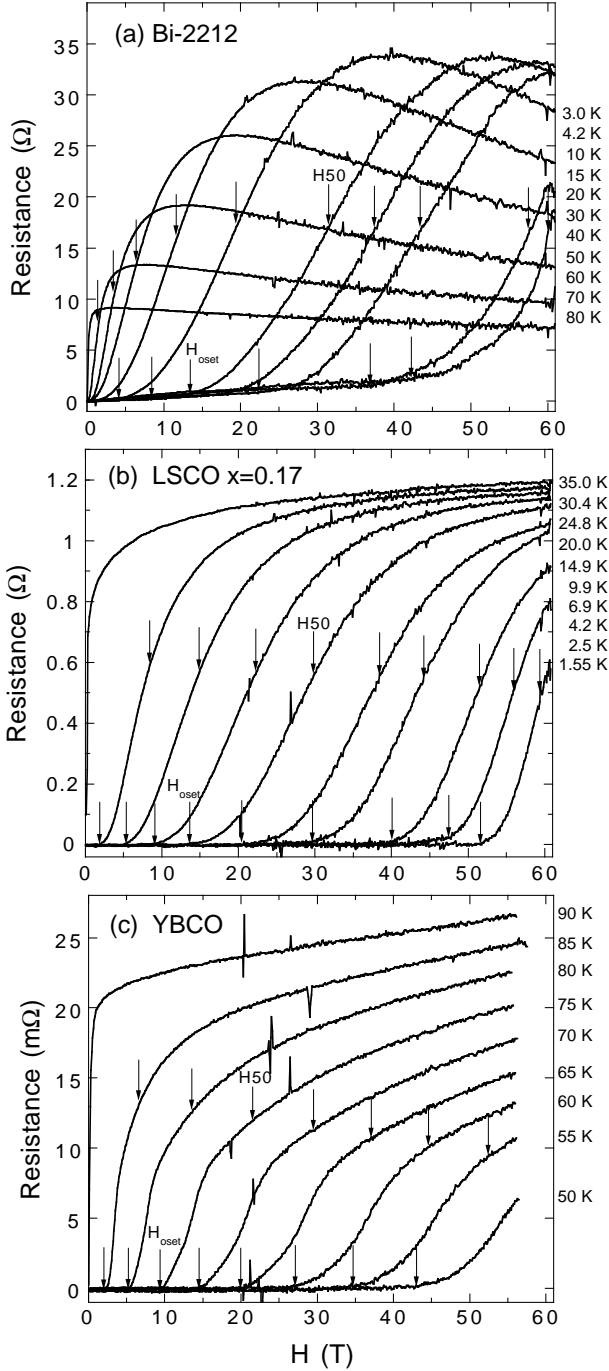}}
\vspace{0.2cm}
\caption{
Selected traces of raw $R$ vs $H$ data for three cuprates near
optimal doping, (a) Bi-2212, (b) LSCO ($x$=0.17), and (c) YBCO.
The temperatures for the traces are listed on the right.
$H_{onset}$ and $H50$ are marked by arrows on the traces.}
\label{fig3}
\end{figure}

Figure 3 contains the raw $R$ vs $H$ data for three cuprates near
optimum doping, Bi-2212, LSCO ($x$=0.17) and YBCO.
One may notice that the data for Bi-2212 [Fig. 3(a)] show some linear 
background before the resistance rapidly increases; 
we have not identified the 
cause of this linear background, but we determined $H_{onset}$ for 
this sample with the deviation from the linear background. 
One may also notice that the Bi-2212 data show pronounced negative
magnetoresistance in the high-field normal state, particularly at 
low temperatures.

In Fig. 3, we show the data in resistance, rather than in resistivity, 
to demonstrate that the noise level of our pulsed-field experiments is 
not determined by the absolute voltage but changes with the sample 
impedance; 
the noise is always about a few percent of the impedance being measured 
(as long as the impedance is larger than a few tens of m${\rm \Omega}$).
This fact precludes the usage of the usual criteria for the 
irreversibility field defined by a certain electric-field threshold.
This is the reason why we needed to determine the onset field $H_{onset}$ 
rather naively with the magnetic field at which the resistivity
first is detected to deviate from zero in the $\rho$ vs $H$ plot.

\begin{figure}
\epsfxsize=0.95\columnwidth
\centerline{\epsffile{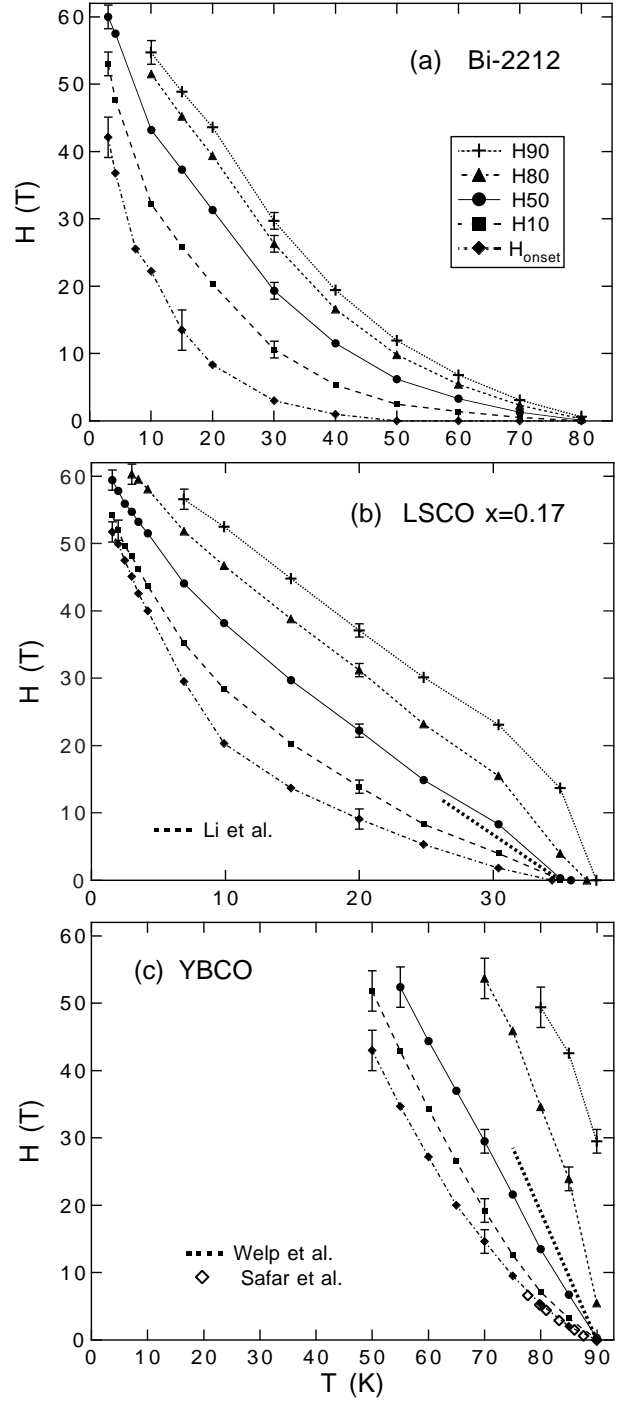}}
\vspace{0.2cm}
\caption{
The contours of the resistive transition in the $H$ vs $T$ plane 
for three cuprates at near optimum doping: 
(a) Bi-2212, (b) LSCO ($x$=0.17), and (c) YBCO.
The thick dashed lines in (b) and (c) are magnetically determined
$H_{c2}$ lines \protect\cite{Li,Welp}. The open diamonds in (c) give
the vortex lattice melting line obtained by Safar {\it et al.}
\protect\cite{Safar}. Note that the panels are arranged
in order of decreasing normal-state anisotropy, $\rho_c/\rho_{ab}$.
Representative error bars are attached to the data.}
\label{fig4}
\end{figure}

In determining the normal-state resistance $R_n$ for the samples shown 
in Fig. 3, we used a linear extrapolation from higher temperatures; 
for example, we estimated $R_n$ to be 33 ${\rm \Omega}$ at 3.0 K for 
Bi-2212, 1.08 ${\rm \Omega}$ at 1.55 K for LSCO ($x$=0.17), and 
14.6 m${\rm \Omega}$ at 50 K for YBCO.
One can infer from Fig. 3 that $R_n$ of Bi-2212 and LSCO (x=0.17) becomes 
relatively temperature independent at low temperatures and therefore the
errors involved in the estimation of $R_n$ are relatively small compared 
to that for YBCO.
The data in Fig. 3(c) show large positive ``magnetoresistance",
particularly at lower temperatures, which makes the definition of 
$R_n$ rather ambiguous.  As a result, the estimated errors for 
$H10$ {\it etc.} are comparatively large for YBCO.
These very broad resistive transitions of YBCO suggest that the idea of 
the ``resistive upper critical field" at which the normal-state
resistivity is restored might be a questionable concept for some
of the high-$T_c$ cuprates.

Figure 4 shows the contours of the resistive transitions in the $H$ vs 
$T$ plane for the three cuprates at near optimum doping, deduced from 
the data in Fig. 3.  Representative error bars are attached to the data.
To corroborate the pulsed-magnetic-field data with more conventional
measurements, we also plot the $H_{c2}(T)$ lines determined from the
reversible magnetization (thick dashed lines): in Fig. 4(b),
for LSCO (x=0.17) crystals from the same source \cite{Li} and, in Fig. 
4(c), the frequently-cited line of -1.9 T/K for YBCO \cite{Welp}.
Note that in each case the $H_{c2}$ line determined from the reversible
magnetization lies near the midpoint of the resistive transition, the
$H50$ line. In Fig. 4(c), the open diamonds denote the first-order
vortex lattice melting transition line obtained by Safar {\it et al.}
\cite{Safar} and this line agrees quite well with our 
$H_{onset}$ (filled diamonds).

The data in Fig. 4 are arranged in order of decreasing normal-state
anisotropy, $\rho_c/\rho_{ab}$. Note that the anomalous upward
curvature, $d^2 H^{\rho}/dT^2 > 0$, is greatest in the more anisotropic
materials. This curvature decreases monotonically, 
until for YBCO the $H80$ and $H90$ curves exhibit
the more conventional negative curvature. We note that this trend does
not depend monotonically on the magnitude of $T_c$, for example.
Furthermore, observation of this trend among different optimally-doped
cuprates suggests that the carrier-concentration dependence found
in LSCO may arise simply from the decreasing normal-state anisotropy
which accompanies increasing carrier concentration.

The $H$ vs $T$ diagrams shown in Figs. 2 and 4 clearly indicate
that, for all the cuprates studied, the unconventional upward curvature
is common for $H_{onset}$ over the entire temperature range studied
here. The anisotropy and short coherence length of the cuprates
give a phase diagram with a vortex liquid regime between the vortex
solid and the normal state\cite{Farrell}. The width of the vortex liquid
regime is expected to increase with increasing anisotropy 
\cite{Tallon}.
It is most likely that our $H_{onset}$ is closely related to the 
irreversibility line, $H_{irr} (T)$, associated with the vortex
liquid-to-solid transition, which would be expected to exhibit the 
observed monotonic dependence on anisotropy.

The interpretation of the rest of the resistive transition,
characterized by the $H10$, $H50$, $H80$ and $H90$ curves of 
Figs. 2 and 4, is not so clear. 
Nonetheless, obvious trends appear in the data. 
For all the cuprates studied, the unconventional upward curvature seen 
in $H_{onset}$ is also seen in $H10$ and $H50$ over the entire 
temperature range studied; however, the upward curvature is observed 
in the higher resistance contours, $H80$ and $H90$, only in such highly 
anisotropic systems as Bi-2212 or underdoped LSCO. 
On the other hand, the lack of saturation of $H^{\rho}_{c2}(T)$ at
low temperatures seems to be a more robust feature of the resistive
transition.
Thus, while there is a substantial literature addressing the resistive
transition and $H^{\rho}_{c2}(T)$ in the cuprates \cite{Brandow}, 
a successful theory would need to account for the following new
phenomenology:
(i) Among the optimally-doped cuprates, $H^{\rho}_{c2}(T)$ can exhibit a
global positive curvature, but only in the more anisotropic compounds.
(ii) As the anisotropy is decreased, the anomalous upward curvature
disappears near $T_c$ in the upper part of the resistive
transition, i.e. in the $H80$ and $H90$ curves. 
(iii) There is no evidence of saturation in $H^{\rho}_{c2}(T)$ in any 
of the cuprates studied in the low temperature limit.

Several dramatically contrasting models for $H_{c2}(T)$ have found 
global positive curvature: for example, localization of charged bosons 
in the small coherence length limit \cite{Alexandrov}, 
scattering from magnetic impurities which order at low temperatures 
\cite{Ovchinnikov}, influence of a quantum critical point associated 
with melting of the vortex lattice \cite{Kotliar},
and mixing of $d_{xy}$ and $d_{x^2-y^2}$ components due to the
magnetic field \cite{Koyama}. 
For many of the above models, the effect of anisotropy is not yet clear.
Our data clearly indicate that
the normal state anisotropy plays a key role, perhaps {\it the} key
role, in determining the curvature of $H^{\rho}_{c2} (T)$.

In summary, we characterize the resistive transition for a variety
of optimally-doped high-$T_c$ cuprates using 61-T pulsed magnetic
fields. The anomalous positive curvature of both $H^{\rho}_{c2} (T)$
and $H^{\rho}_{irr} (T)$ is strongest in the more anisotropic
cuprates. None of the samples studied show saturation of 
$H^{\rho}_{c2} (T)$ or $H^{\rho}_{irr} (T)$ at low temperatures.

The authors would like to acknowledge helpful discussions with D.J.
Bishop, J.R. Cooper, D.A. Huse, R. Ikeda, A.N. Lavrov, A.P. Mackenzie, 
A. Sudbo, and C.M. Varma.

% Place here the list of the references:
%


\begin{references}

\bibitem{Fisher2huse}D.S. Fisher, M.P.A. Fisher, and D.A. Huse, Phys.
Rev. {\bf B43}, 130 (1991).

\bibitem{Safar2}H.Safar, P.L. Gammel, D.A. Huse, D.J. Bishop, W.C. Lee,
J. Giapintzakis, and D.M. Ginsberg, Phys. Rev. Lett. {\bf 70}, 3800
(1993).

\bibitem{Farrell}
For review, see D.E. Farrell, in {\it Physical Properties of High
Temperature Superconductors IV}, edited by D.M. Ginsberg (World
Scientific, Singapore, 1994).

\bibitem{Mackenzie}
A. P. Mackenzie, S. R. Julian, G. G. Lonzarich, A. Carrington,
S. D. Hughes, R. S. Liu, and D. C. Sinclair, Phys. Rev. Lett.
{\bf 71}, 1238 (1993).

\bibitem{Osofsky}
M. S. Osofsky, R. J. Soulen, Jr., S. A. Wolf, J. M. Broto,
H. Rakoto, J. C. Ousset, G. Coffe, S. Askenazy, P. Pari, I. Bozovic,
J. N. Eckstein, and G. F. Virshup, Phys. Rev. Lett. {\bf 71}, 2315
(1993).

\bibitem{WHH}
N.R. Werthamer, E. Helfand, and P.C. Hohenberg, Phys. Rev. {\bf 147},
295 (1966).

\bibitem{Brandow}For a review, see B. Brandow, Phys.
Rep. {\bf 296}, 1 (1998).

\bibitem{LosAlamos}
An exception is an experiment using explosively-generated microsecond
magnetic fields [Smith {\it et al.}, J. Superconductivity {\bf 7}, 269
(1994)],
suggesting that $H^{\rho}_{c2} (T)$ of optimally-doped YBCO
is consistent with WHH; however, $H^{\rho}_{c2} (T)$ in these difficult
measurements is defined at a given fixed $T< T_c$, not by any observed
feature in the $\rho (B)$ data, but rather by the magnetic field at
which the measured $\rho (B)$ equals the normal state resistivity
at $T=T_c$.

\bibitem{Karpinska} K. Karpinska, A. Malinowski, Marta Z. Cieplak, S.
Guha, S. Gershman,
G. Kotliar, T. Skoskiewicz, W. Plesiewicz, M. Berkowski, and P.
Lindenfeld, Phys. Rev. Lett. {\bf 77}, 3033 (1997).

\bibitem{Walker}D.J.C. Walker, O. Laborde, A.P. Mackenzie, S.R. Julian,
A. Carrington,
J.W. Loram, and J.R. Cooper, Phys. Rev. {\bf B51}, 9375 (1995).

\bibitem{Dalichaouch}Y. Dalichaouch, B.W. Lee, C.L. Seaman, J.T.
Markert, and M.B. Maple, Phys. Rev. Lett. {\bf 64}, 599 (1990).

\bibitem{Carrington}A. Carrington, A.P. Mackenzie, and A. Tyler, Phys.
Rev. {\bf B54}, R3788 (1996).

\bibitem{Blumberg}G. Blumberg, Moonsoo Kang, and M.V. Klein, Phys. Rev.
Lett. {\bf 78}, 2461 (1997).

\bibitem{Bergemann}
C. Bergemann, A.W. Tyler, A.P. Mackenzie, J.R. Cooper, S.R. Julian, 
and D.E. Farrell, Phys. Rev. B {\bf 57}, 14387 (1998).

\bibitem{x-dep-PRL}
G.S. Boebinger, Y. Ando, A. Passner, K. Tamasaku, N. Ichikawa, S.
Uchida, M. Okuya, T. Kimura, J. Shimoyama, and K. Kishio,
Phys. Rev. Lett. {\bf 77}, 5417 (1996).

\bibitem{Bi2201-PRL}
Y. Ando, G.S. Boebinger, A. Passner, N. L. Wang, C. Geibel, and F.
Steglich,
Phys. Rev. Lett. {\bf 77}, 2065 (1996); {\bf 79}, 2595(E) (1997).

\bibitem{logT-PRL}
Y. Ando, G.S. Boebinger, A. Passner, T. Kimura, and K. Kishio,
Phys. Rev. Lett. {\bf 75}, 4662 (1995).

\bibitem{Li}
Q. Li, M. Suenaga, T. Kimura, and K. Kishio, Phys. Rev. B {\bf 47},
11384 (1993).

\bibitem{Welp}
U. Welp, W.K. Kwok, G.W. Crabtree, K.G. Vandervoort, and J.Z. Liu,
Phys. Rev. Lett. {\bf 62}, 1908 (1989).

\bibitem{Safar}
H. Safar, P.L. Gammel, D.A. Huse, D.J. Bishop, J.P. Rice, and D.M.
Ginsberg,
Phys. Rev. Lett. {\bf 69}, 824 (1992).

\bibitem{Tallon}
J.L. Tallon, G.V.M. Williams, C. Bernhard, D.M. Pooke, M.P. Staines, 
J.D. Johnson, and R.H. Meinhold, Phys. Rev. B {\bf 53}, R11972 (1996).

\bibitem{Alexandrov}
A.S. Alexandrov, Phys. Rev. {\bf B48}, 10571 (1993) and A.S. Alexandrov,
V.N. Zavaritsky,
W.Y. Liang, and P.L. Nevsky, Phys. Rev. Lett. {\bf 76}, 983 (1996).

\bibitem{Ovchinnikov}
Y. Ovchinnikov and V. Kresin, Phys. Rev. {\bf B52}, 3075 (1995).

\bibitem{Kotliar}
G. Kotliar and C.M. Varma, Phys. Rev. Lett. {\bf 77}, 2296 (1996).

\bibitem{Koyama}
T. Koyama and M. Tachiki, Physica (Amsterdam) C {\bf 263}, 25 (1996).

\end{references}
\end{document}